\documentclass[11pt]{article}
\usepackage[sort&compress,square,comma,numbers]{natbib}

\usepackage{textcomp}
\usepackage{graphicx}

\usepackage[labelfont=bf]{caption} 
\usepackage[font={small}]{caption}   

\usepackage{subcaption}
\usepackage{amsmath}
\usepackage{amssymb}
\usepackage{url}
\usepackage{placeins}

\usepackage[usenames,dvipsnames]{color}    

\usepackage{soul}   
\setstcolor{red} 
\setul{}{1.5pt}  

\newcommand{\xiv}{\ensuremath{\hat\xi} }
\newcommand{\xivLO}{\ensuremath{\hat\xi_\textrm{LO}} }
\newcommand{\xivTO}{\ensuremath{\hat\xi_\textrm{TO}} }
\newcommand{\Rxiv}{\ensuremath{R_{\hat\xi}} }
\newcommand{\xv}{\ensuremath{\hat x}}
\newcommand{\yv}{\ensuremath{\hat y}}
\newcommand{\zv}{\ensuremath{\hat z}}
\newcommand{\Rxv}{\ensuremath{R_{\hat x}}}
\newcommand{\Ryv}{\ensuremath{R_{\hat y}}}
\newcommand{\Rzv}{\ensuremath{R_{\hat z}}}
\newcommand{\kv}[1]{\ensuremath{\vec{k_{#1}}}}
\newcommand{\qv}{\ensuremath{\vec{q}} }
\newcommand{\ev}[1]{\ensuremath{\hat{e_{#1}} }}

\title{\textbf{Raman Cooling of Solids through Photonic Density of States Engineering}}
\author{Yin-Chung Chen$^1$, Gaurav Bahl$^{1\ast}$\\
	\\
	\footnotesize{$^1$Mechanical Science and Engineering, University of Illinois at Urbana-Champaign,}\\
	\footnotesize{Urbana, Illinois, USA}\\
	\footnotesize{$^\ast$To whom correspondence should be addressed; E-mail: bahl@illinois.edu.}
	}
\date{}
\begin{document}

\maketitle

\begin{abstract}
The laser cooling of vibrational states of \textit{solids} has been achieved through photoluminescence in rare-earth elements, optical forces in optomechanics, and the Brillouin scattering light-sound interaction. 
The net cooling of solids through spontaneous Raman scattering, and laser refrigeration of indirect band gap semiconductors, both remain unsolved challenges.
Here, we analytically show that photonic density of states (DoS) engineering can address the two fundamental requirements for achieving spontaneous Raman cooling: suppressing the dominance of Stokes (heating) transitions, and the enhancement of anti-Stokes (cooling) efficiency beyond the natural optical absorption of the material.
We develop a general model for the DoS modification to spontaneous Raman scattering probabilities, and elucidate the necessary and minimum condition required for achieving net Raman cooling.
With a suitably engineered DoS, we establish the enticing possibility of refrigeration of intrinsic silicon by annihilating phonons from all its Raman-active modes simultaneously, through a single telecom wavelength pump.
This result points to a highly flexible approach for laser cooling of any transparent semiconductor, including indirect band gap semiconductors, far away from significant optical absorption, band-edge states, excitons, or atomic resonances.
\end{abstract}

\section{Introduction}

The photon-induced annihilation of thermal quanta from matter, i.e. laser cooling, is of key importance in ultracold science and has been instrumental in the creation of gas phase quantum condensates, matter-wave interferometry, and macroscopic tests of quantum mechanics.
The laser cooling of solids, in particular, is enabled through photon up-conversion processes in which phonons are annihilated from the material \cite{pringsheim1929zwei, PhysRevLett.92.247403, ding2012anti}. On this principle, bulk cooling of solids has been achieved through photoluminescence on specific electronic transitions in Yb-doped solids \cite{epstein1995observation} with demonstrations even reaching cryogenic temperatures \cite{seletskiy2010laser, seletskiy2012cryogenic, melgaard2013optical, melgaard2014identification}. 
Photoluminescence cooling, enhanced through exciton-phonon coupling, was also employed for the first demonstration of laser cooling of an undoped direct band gap semiconductor \cite{zhang2013laser}. However, this method is not effective with \textit{indirect} band gap semiconductors like silicon and germanium due to the need of a phonon for momentum conservation and the consequent competition with nonradiative recombination.
In parallel efforts, efficient narrowband cooling of individual phonon modes in solids has also been achieved by means of optical forces in geometry-engineered microstructures \cite{arcizet2006radiation, gigan2006self, kleckner2006sub} and through Brillouin scattering \cite{bahl2012observation}, albeit without net cooling.
The search still continues for a universal laser refrigeration mechanism that could be applied to any transparent material, with any geometry, at any optical wavelength.

In this context, Raman scattering of light from optical phonons in solids has been suggested as a promising alternative technology, especially since Raman scattering is available in all materials and also exhibits the requisite photon up-conversion i.e. anti-Stokes scattering \cite{bahl2012observation, rand2013raman, ding2012anti}. 
The key distinction is that Raman cooling involves only virtual excited states while photoluminescence mechanisms involve real excitations of the electronic states of a system. Photoluminescence thus relies on the existence of suitable natural materials, while Raman scattering can be potentially engineered.
However, two key challenges must be resolved before Raman cooling of a bulk solid can be achieved. 
First, spontaneous Stokes Raman scattering always dominates over anti-Stokes scattering in bulk materials due to the event probabilities being scaled by $n_0+1$ and $n_0$ respectively, where $n_0=(e^{\hbar\omega_0/k_B T}-1)^{-1}$ is the Bose-Einstein distribution function. It is essential to invert this imbalance to have any possibility of achieving cooling.
Second, the anti-Stokes scattering efficiency needs to be enhanced so as to overcome absorption of the pump light by the material. This is critical since each anti-Stokes Raman scattering event annihilates $\hbar\omega_0$ energy from the material ($\omega_0$ is phonon frequency), while each absorption event adds $\hbar\omega_\textrm{photon} (\gg \hbar\omega_0)$ energy to the material.

Here we show that both of the above requirements can be met in nearly any semiconductor by engineering the optical density of states to a suitable form. 
Accordingly, we develop expressions for the spontaneous Raman scattering efficiency in a material with a modified photonic density of states (DoS), and derive the necessary and minimum condition to achieve cooling by this method.
As a specific case, we demonstrate the surprising possibility of cooling all the Raman-active phonon modes in undoped crystalline silicon using a single telecom wavelength pump.

\vspace{9pt}

Before we proceed, we must recognize the encouraging experimental efforts of past researchers also moving in the direction of this result.
It has been reported that resonant Raman scattering (using excitonic enhancement) in certain semiconductors \cite{ding2012anti, zhang2013laser} can be used to tilt the Stokes vs anti-Stokes balance in favor of cooling. 
Ding and Khurgin showed \cite{ding2012anti} that by pumping GaN at 375.8 nm, anti-Stokes to Stokes ratio of 0.38 can be naturally reached. More recently, Zhang et al reported \cite{zhang2013laser} a ratio of 1.43 in CdS nanobelts by means of an exciton resonance, firmly establishing that anti-Stokes dominance can indeed be achieved.
Since this method relies on interaction with an exciton, it can only allow narrowband cooling of one or a few phonon modes that lie within the resonance, a constraint similar to optomechanical cooling \cite{arcizet2006radiation, gigan2006self, kleckner2006sub, bahl2012observation}. 
Additional challenges to this method are the large optical absorption that occurs due to strong exciton-photon interaction, and its impracticality for cooling indirect band gap semiconductors. 
A different perspective on Raman cooling has been offered by Rand \cite{rand2013raman}, using a multi-pump method on specific electronic transitions of rare earth elements like Ce$^{3+}$ and Yb$^{3+}$. However, as in other efforts each photon pumping configuration chosen would target only one phonon mode.

\begin{figure}[h!]
	\begin{center}
		\includegraphics[trim=0cm 2.5cm 0cm 1cm, clip, width=0.7\textwidth]{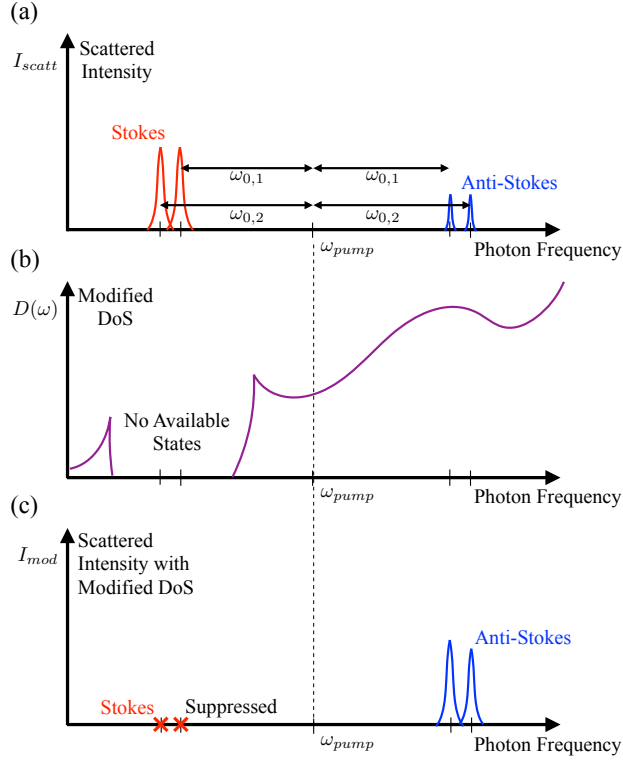}
		\caption{Concept of density of states (DoS) engineering for Raman cooling of an arbitrary number of phonon modes.
		(a) Stokes scattering (red) dominates anti-Stokes scattering (blue) from any chosen phonon mode $\omega_{0,i}$ in bulk media. 
		(b) It is proposed that an engineered photonic DoS with a complete band gap can be used to (c) suppress Stokes scattering while simultaneously enhancing anti-Stokes scattering intensity, as previously demonstrated with Brillouin cooling \cite{bahl2012observation}.
	}
	\label{fig:RS}
	\end{center}
\end{figure}

We propose that engineering the photonic 
DoS \cite{purcell1946spontaneous, gaponenko2002effects} at high-transparency wavelengths can be used to invert the imbalance of Stokes vs anti-Stokes spontaneous Raman scattering efficiencies, and can also be used to enhance anti-Stokes Raman scattering through engineered photonic resonances. 
This proposal is outlined in Fig.~\ref{fig:RS}.
This method circumvents the optical absorption issue confronted by techniques that use excitons \cite{ding2012anti, zhang2013laser}, can be applied to indirect band gap semiconductors, and also avoids band-tail absorption confronted in photoluminescence approaches \cite{khurgin2006band}.
Such DoS engineering has been previously employed in experiments with W1 silicon waveguides \cite{checoury2009enhanced} to enhance spontaneous Stokes Raman scattering by a factor up to 13.
It has also been shown that the naturally occurring photonic DoS of a high-Q resonator can be used to enhance anti-Stokes Brillouin scattering by orders-of-magnitude while completely suppressing Stokes scattering, thus enabling Brillouin cooling \cite{bahl2012observation}. 
The possibility of by-design DoS engineering to achieve spontaneous anti-Stokes Raman cooling is clear.

\section{Raman scattering efficiency with modified DoS}
Consider a spontaneous Raman scattering process with incident photon frequency $\omega_i$ and scattered photon frequency $\omega$. The scattering probability is given by \cite{gaponenko2002effects, sushchinskiui1972raman}
\begin{equation}\label{eq:SP}
W(\omega_i, \omega)
	=\frac{2\pi^2}{\hbar^2}
	\omega_i\omega N_i\frac{1}{4\pi}
	D(\omega)
	|\mathcal{M}|^2,
\end{equation}
where $N_i$ is the number of incident photons, $\hbar$ is the Planck constant, $D(\omega)$ is the photonic DoS at the scattered photon frequency, and $\mathcal{M}$ is the matrix element of the scattering process. Here the scattered photon frequency $\omega$ can be taken as either the Stokes frequency $\omega_S=\omega_i-\omega_0$ or the anti-Stokes frequency, $\omega_{AS}=\omega_i+\omega_0$, where $\omega_0$ is the phonon frequency. If the scattering process happens in free space, the DoS is obtained using the linear dispersion relation $\omega = c | \vec{k} |$.
\begin{equation}\label{eq:DoS3D}
D_3(\omega)
	=\frac{\omega^2}{2\pi^2c^3},
\end{equation}
where \kv~ is the wave vector of the photon and $c$ the speed of light. If we assume the phonon energy is small compared against the photon energy, we can approximate $\omega_i\approx\omega$, and Eqn. (\ref{eq:SP}) will recover the well-known $W\propto\omega^4$ relation \cite{gaponenko2002effects}. On the other hand, in a medium where the photon dispersion relation is not given by $\omega = c | \vec{k} |$, the scattering probability will be obtained by replacing the photonic DoS in Eqn.~\ref{eq:DoS3D} by the modified DoS. In such a generalized case, the scattering efficiency per unit length, $\delta S$, per unit solid angle, $\delta\Omega$, is given by
\begin{equation}
\frac{\partial S}{\partial\Omega}
	\propto\frac{W\hbar\omega}{N_i\hbar\omega_i}
	\propto D(\omega).
\end{equation}
The major challenge in calculating the Raman scattering efficiency is to determine the matrix element $\mathcal{M}$ in Eqn.~\ref{eq:SP}.

To quantitatively evaluate the effects of arbitrary photonic DoS on the Raman scattering efficiency, we use the formulae found in \cite{wagner1983absolute, aggarwal2011measurement} for the following calculations. The Raman scattering efficiency per unit length and solid angle is given by \cite{wagner1983absolute} as
\begin{equation}\label{eq:dSdOS}
\left (\frac{\partial S}{\partial\Omega}  \right )_{Stokes}
	=\left (\frac{\omega_S}{c} \right )^4
	\frac{N\hbar}{M\omega_0}(1+n_0)
	|R_S(\Omega)|^2
\end{equation}
for Stokes scattering, and
\begin{equation}\label{eq:dSdOAS}
\left (\frac{\partial S}{\partial\Omega}  \right )_{Anti-Stokes}
	=\left (\frac{\omega_{AS}}{c} \right )^4
	\frac{N\hbar}{M\omega_0}n_0
	|R_{AS}(\Omega)|^2
\end{equation}
for anti-Stokes scattering, where $N$ is the number of unit cells per unit volume, $M$ is the atomic mass, and $n_0=(e^{\hbar\omega_0/k_B T}-1)^{-1}$ is the phonon occupation number. $R$ is the Raman tensor element, which depends on the scattering angle and the crystal structure of the material. 
In the simple case where the medium is transparent with respect to the pump and scattered light, the total Raman scattering efficiencies per unit length, $S_S$ and $S_{AS}$, can be calculated \cite{yu1996fundamentals} by integrating Eqns. \ref{eq:dSdOS} and \ref{eq:dSdOAS} over the solid angle,
\begin{equation}\label{eq:SS3D}
S_{S}
	=\left (\frac{\omega_S}{c} \right )^4\frac{N\hbar}{M\omega_0}(1+n_0)
	\int_{\Omega}|R_S(\Omega)|^2d\Omega, 
\end{equation}
\begin{equation}\label{eq:SAS3D}
S_{AS}
	=\left (\frac{\omega_{AS}}{c} \right )^4\frac{N\hbar}{M\omega_0}n_0
	\int_{\Omega}|R_{AS}(\Omega)|^2d\Omega. 
\end{equation}
Note that the above equations hold only for linear dispersion relations. Following Eqn.~\ref{eq:SP}, if the photonic DoS is modified Eqns. \ref{eq:SS3D} and \ref{eq:SAS3D} become 
\begin{equation}\label{eq:SS}
S_{S}
	=\left (\frac{\omega_S}{c} \right )^4
	\frac{N\hbar}{M\omega_0}(1+n_0)
	\int_{\Omega}\frac{D(\omega_{S},\Omega)}{D_3(\omega_{S})/4\pi}
	|R_S(\Omega)|^2d\Omega, 
\end{equation}
\begin{equation}\label{eq:SAS}
S_{AS}
	=\left (\frac{\omega_{AS}}{c} \right )^4
	\frac{N\hbar}{M\omega_0}n_0
	\int_{\Omega}\frac{D(\omega_{AS},\Omega)}{D_3(\omega_{AS})/4\pi}
	|R_{AS}(\Omega)|^2d\Omega,
\end{equation}
where $D(\omega, \Omega)$ is the modified DoS at certain solid angle and $D_3(\omega)$ is obtained from Eqn.~\ref{eq:DoS3D}. This DoS ratio factor resembles the Purcell enhancement of spontaneous emission \cite{purcell1946spontaneous}. Note that if the modified DoS is isotropic, $D(\omega, \Omega)$ can be replaced by $D(\omega)/4\pi$ simplifying the factor within the integrals to $D(\omega)/D_3(\omega)$, which is independent of the scattering angle.

Since the typical phonon energies are small compared against the photon energy, the Raman tensor for both processes is symmetrical and has approximately the same values. From Eqns.~\ref{eq:SS} and \ref{eq:SAS}, we see that it is also possible to 
adjust the scattering efficiency in the Stokes and anti-Stokes directions by aligning the material's crystal (Raman selection rules) suitably with respect to the anisotropic photonic DoS.

\section{Application to laser cooling}

{Now that we have obtained the DoS-modified Raman scattering efficiencies, we can calculate whether this can be used to achieve Raman cooling in solids. }
We consider a highly transparent medium 
in which Raman scattering occurs over a volume of {cross-sectional} area $A$ and {effective} length $L$. We can define net scattering efficiencies by $\eta_S = L \, S_S$ and $\eta_{AS} = L \, S_{AS}$, such that $I_{S} = \eta_{S} \, I_{pump}$ and $I_{AS} = \eta_{AS} \, I_{pump}$ are the scattered intensities, with $I_{pump}$ being the incident light intensity. 
The experiment should be performed at high-transparency wavelengths far away from any electronic resonances, exciton resonances, or two photon absorption in the material. It is then fair to assume that the only energy exchange mechanisms between the medium and the pump laser are residual absorption and Raman scattering. Under these assumptions, the net power transferred into the medium is given by
\begin{equation}\label{eq:EB1}
	P_{net}=P_{abs}+P_{ph,S}-P_{ph,AS},
\end{equation}
where $P_{abs}$ is the broad-spectrum absorbed optical power, while $P_{ph,S}$ and $P_{ph,AS}$ are the heating and cooling power due to phonon creation and annihilation from Stokes and anti-Stokes processes, respectively. Cooling of the entire material will occur when $P_{net}<0$.

The absorbed power can be calculated by $P_{abs} = A \, I_{pump} ( 1 - e^{-\alpha L})$, where $\alpha$ is the absorption coefficient at the incident frequency. Since the absorption coefficient is small for a transparent material the absorbed power is approximately $P_{abs}=A \, L \, \alpha \, I_{pump}$. Presently, we consider only one phonon mode for which the scattered optical power in the Stokes sideband is $P_S =A \, \eta_S \, I_{pump}=A \, L \, S_S \, I_{pump}$, and similarly $P_{AS}=A \, L\, S_{AS} \, I_{pump}$ for the anti-Stokes sideband. For each Raman scattered photon, one phonon is added or removed from the system. The heating (cooling) power due to scattering is thus obtained by multiplying the creation (annihilation) rate by the phonon energy $\hbar\omega_0$.
\begin{equation}
P_{ph,S}
	=\hbar\omega_0\frac{P_S}{\hbar\omega_S}
	=A\cdot L\cdot \frac{\omega_0}{\omega_S}S_SI_{pump},
\end{equation}
\begin{equation}
P_{ph,AS}
	=\hbar\omega_0\frac{P_{AS}}{\hbar\omega_{AS}}
	=A\cdot L\cdot \frac{\omega_0}{\omega_{AS}}S_{AS}I_{pump}.
\end{equation}
Upon substituting these relations into equation (\ref{eq:EB1}), we have the cooling condition
\begin{equation}
P_{net}
	=A\cdot L\cdot I_{pump} \left (\alpha+\frac{\omega_0}{\omega_S}S_S
	-\frac{\omega_0}{\omega_{AS}}S_{AS} \right )<0,
\end{equation}
resulting in the necessary and minimum requirement for cooling of the solid
\begin{equation}\label{eq:EBF}
\boxed{\alpha
	<\frac{\omega_0}{\omega_{AS}}S_{AS}
	-\frac{\omega_0}{\omega_S}S_S.
}
\end{equation}
This means that to achieve cooling, the net cooling efficiency per unit length $\left( \frac{\omega_0}{\omega_{AS}}S_{AS} - \frac{\omega_0}{\omega_S}S_S \right)$ must exceed the optical absorption coefficient. If the DoS is designed such that multiple Raman modes are simultaneously affected, the above relation can be suitably expanded for multiple phonon modes into the form
\begin{equation}
	\label{eq:EBF2}
	\alpha
		<	\sum_{\textrm{all modes}}	 \left(  \frac{\omega_0}{\omega_{AS}} S_{AS}
		- \frac{\omega_0}{\omega_S} S_S   \right).
\end{equation}

%
%
\section{Raman cooling of silicon}

Amongst common semiconductor materials, {intrinsic crystalline} silicon is extremely transparent in the telecom range \cite{green1995optical, green2008self, keevers1995absorption}, and has Raman scattering {efficiency} comparable to the optical absorption {coefficient} \cite{aggarwal2011measurement}.
It is thus a promising material with which to demonstrate Raman cooling.
In this section, we show that the cooling of undoped silicon is practical using a telecom pump when a diamond structure 3D photonic crystal is used to generate a complete photonic band gap.

First, we consider a simplification to Eqns. \ref{eq:SS} and \ref{eq:SAS} for cases where an isotropic or nearly isotropic photonic DoS is available, {which will be used later.} The ratio of DoS in these equations can then be pulled out from the integral as a simple prefactor $D(\omega)/(\frac{\omega^2}{2\pi^2c^3})$, where the numerator is the photonic DoS of the crystal, and the denominator is the vacuum DoS.
We then simply need to integrate the variation of the Raman tensor element $\left( \int_{\Omega}|R|^2 d \Omega \right)$ over the solid angle $\Omega$ over which scattering takes place. 
In the case of crystalline silicon (point group $O_h$), we can define 
\xv, \yv, and \zv \:  
as the [100], [010], and [001] directions of the material crystal. The Raman tensor for phonon polarization in each of these directions is \cite{loudon1964raman, yu1996fundamentals}
\begin{equation}\label{eq:RT}
	\Rxv = \begin{pmatrix}
		0 & 0 & 0\\
		0 & 0 & d\\
		0 & d & 0\\
	\end{pmatrix},\ 
	\Ryv = \begin{pmatrix}
		0 & 0 & d\\
		0 & 0 & 0\\
		d & 0 & 0\\
	\end{pmatrix},\ 
	\Rzv = \begin{pmatrix}
		0 & d & 0\\
		d & 0 & 0\\
		0 & 0 & 0\\
	\end{pmatrix}, 
\end{equation}
where $d$ is the only independent Raman tensor element for crystalline silicon since the three optical phonon modes in the $\Gamma$ point are triply degenerate \cite{yu1996fundamentals}. {The triply degenerate feature of the optical phonons manifests itself in the common Raman tensor element $d$ for both LO and TO modes}. Note that for a polar material the Raman tensor element $d$ for longitudinal optical (LO) and transverse optical (TO) phonons will be different due to elasto-optic effects.
Let us further consider an incident photon with polarization in vector direction 
\ev{i},
scattered into polarization 
\ev{s}
in this crystal. The Raman tensor element in Eqns. \ref{eq:SS} and \ref{eq:SAS} is given by
\begin{equation}\label{eq:R2}
	|R|^2
	= | \ev{i} \cdot \Rxiv \cdot \ev{s} |^2,
\end{equation}
where 
\xiv
is the phonon polarization unit vector, and 
\Rxiv
is the corresponding linear combination of the 
\Rxv, \Ryv, and \Rzv \:
matrices \cite{yu1996fundamentals}. 
In our calculation we fix the incident photon to be traveling in the direction $\kv{i} = \xv$, 
having polarization $\ev{i} = \zv$
with respect to the material crystal. We then quantify light scattering for all possible scattered momentum vectors \kv{s}.
For each \kv{s} 
two mutually perpendicular photon polarizations (\ev{s})
are considered.
Conservation of momentum enforces the relationship 
$\kv{i} \pm \qv = \kv{s}$
where 
\qv
is the phonon wave vector.
For each 
\qv
we must consider three phonon polarizations, composed of one \xivLO
parallel to 
\qv
and two 
\xivTO
orthogonal to 
\qv.
\Rxiv is evaluated based on these choices, and the value of the Raman tensor element $|R|^2$ can be computed.
After summing for the two possible scattered photon polarizations, we obtain the total contributions from LO mode and two TO modes to the integral $\int_{\Omega}|R|^2d\Omega$ as $6.28 |d|^2$ and $10.47 |d|^2$, respectively. {These values are valid for all materials in the same crystal structure with this particular scattering geometry (\xv ~incident light with \zv ~polarization).}
{For silicon, since $d$ for all three degenerate modes are the same, we collect all the contributions in to one single scattering efficiency.} This yields a total value of $16.76 |d|^2$ to the scattering efficiency {from all three phonon modes} in an isotropic DoS medium.

\vspace{9pt}

\begin{figure}[tb]
	\begin{center}
	\includegraphics[trim=0.5cm 3cm 0.5cm 1.6cm, clip, width=0.8\textwidth]{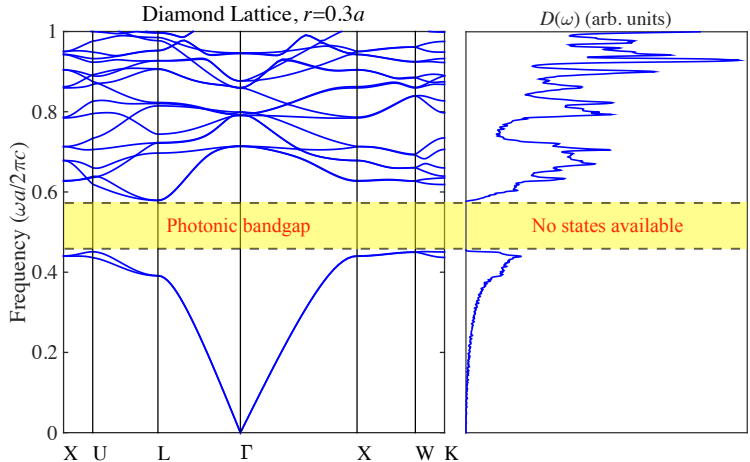}
	\caption{Band structure and DoS of a diamond structure photonic crystal consisting of air spheres in silicon. The refractive index of silicon is $n=3.44$ in the calculation, the radius of the air-spheres is $r=0.3a$, where $a$ is the lattice constant of the photonic crystal. The yellow shaded region denotes the range of the photonic band gap. The frequency is in non-dimensional units $\omega a /2\pi c$. \\
	}
	\label{fig:DPC}
	\end{center}
\end{figure}

\begin{figure}[t!]
	\centering
		\includegraphics[clip=true, trim=2.0cm 0cm 3cm 0cm, width=0.6\textwidth]{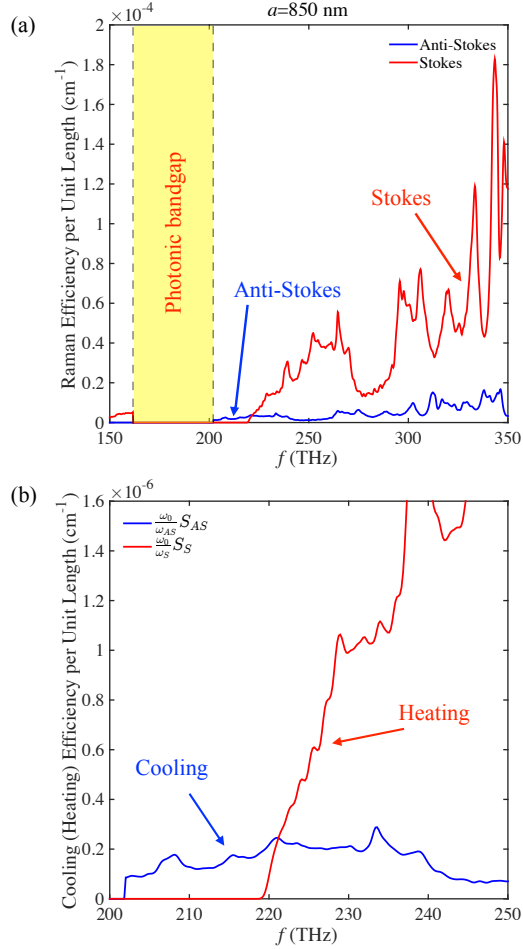}
		\caption{
		\textbf{Cooling all Raman-active phonons in silicon:}
			(a) Raman efficiencies per unit length $S_{AS}$ and $S_{S}$ (in intrinsic crystalline silicon patterned with the Fig.~\ref{fig:DPC} diamond photonic crystal) are functions of pump frequency. The contributions from all three Raman-active phonon modes are included. The yellow shaded region denotes the band gap for the pump.
		Band-edge absorption, not included, will become significant at pump frequency higher than 260 THz ($\sim$ band gap energy).
		(b) Calculated cooling and heating efficiency per unit length.
		The Stokes process is suppressed over a wide range resulting in net phonon energy removal and the simultaneous cooling of all Raman-active phonons.
				}
		\label{fig:Eff}
\end{figure}

The net cooling condition (Eqn. \ref{eq:EBF}), demands minimization of the Stokes efficiency, thus a photonic DoS design that can fully suppress Stokes scattering is strongly desired. This is available in various three dimensional photonic crystal structures, including diamond lattice \cite{ho1990existence}, Yablonovite \cite{yablonovitch1991photonic}, woodpile crystal \cite{ho1994photonic, sozuer1994photonic}, inverse-opals \cite{vlasov2001chip}, and two dimensional crystal stacks \cite{joannopoulos2011photonic6}. For this study, we select spherical air inclusions in silicon, structured as a diamond lattice.

In Fig.~\ref{fig:DPC} we present the well-known photonic band structure and DoS of a diamond structure silicon-air photonic crystal, with radius of air spheres $r=0.3a$, where $a$ is the lattice constant. These calculations are performed using MIT Photonic-Bands package \cite{Johnson2001:mpb}. Since the first Brillouin zone of a diamond structure (face centered cubic) crystal is a truncated octahedron, which is nearly spherical \cite{joannopoulos2011photonic6}, we can make the simplifying assumption that the photonic DoS is approximately isotropic. We can then compute the photonic DoS through \cite{Ashcroft}
\begin{equation}\label{eq:DoS}
D(\omega)
	=\sum_{i} \int_{\textrm{FBZ}} \frac{d \kv{} }{(2\pi)^3}
	~ \delta \left( \omega-\omega_i( \kv{} ) \right),
\end{equation}
where $i$ is the band index, and the integral is over the first Brillouin zone (FBZ) of the photonic crystal.

We now impose this DoS onto the Raman efficiency formulae and assume that the escape efficiency for the scattered photons is 1, i.e. all the scattered light escapes the medium \cite{sheik2009laser}.
In Figure \ref{fig:Eff}a we plot the total scattering efficiency per unit length of the Stokes ($S_S$) and anti-Stokes ($S_{AS}$) processes as functions of pump frequency for the selected photonic crystal. Note that we include the contribution of the triply degenerate LO/TO phonon modes of silicon into a single $S_S$ and $S_{AS}$. We invoke the known properties of silicon \cite{aggarwal2011measurement}, $M=28.09$ amu, $N=2.5\times 10^{22}$ cm$^{-3}$, $|d|=1.9\times 10^{-15}$ cm$^2$, and $a=850$ nm as the lattice constant of the diamond structure photonic crystal to complete our calculations. 
%
%
Several notes can be made from this computation.
Pump light lies within the photonic band gap in the 160 -- 203 THz (1476 -- 1873 nm) range. Therefore, little to no scattering will occur. Pumping in the 203 -- 220 THz (1362 -- 1476 nm) range, however, completely suppresses the Stokes process. Simultaneously, the anti-Stokes process is enhanced by a factor of $\sim 15$.

\vspace{9pt}

\begin{figure}[tb]
	\begin{center}
		\includegraphics[trim=4.0cm 0.9cm 4.3cm 1.1cm, clip, width=0.6\textwidth]{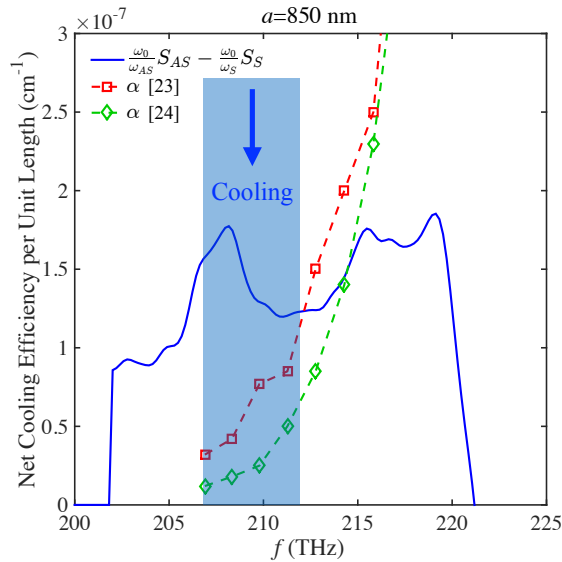}
		\caption{
		\textbf{Net Raman cooling is achievable in silicon:}
		The net cooling efficiency per unit length ($\frac{\omega_0}{\omega_{AS}}S_{AS}-\frac{\omega_0}{\omega_S}S_S$) can exceed the absorption coefficient for the design presented. The absorption coefficient for silicon is extracted from \cite{green1995optical, green2008self}. Near 210 THz (1427 nm), the net cooling over comes the absorption.
		}
		\label{fig:C}
	\end{center}
\end{figure}

In Fig.~\ref{fig:Eff}b, we plot the cooling and heating efficiencies per unit length, i.e. $\frac{\omega_0}{\omega_{AS}}S_{AS}$  and $\frac{\omega_0}{\omega_{S}}S_{S}$ as a function of pump frequency. 
We observe that net cooling can occur within the Stokes suppression range (203-220 THz, 1362 nm - 1476 nm) as long as Eqn.~\ref{eq:EBF} can be satisfied. 
The majority of optical absorption in crystalline silicon (band gap energy $\approx 1.12$ eV) over this frequency range comes from {three- and four-phonon assisted absorption} \cite{keevers1995absorption}, and is very small. 
To evaluate the possibility of cooling, we invoke experimentally measured absorption coefficients from \cite{green1995optical, green2008self} and compare them against the net cooling efficiency per unit length $\frac{\omega_0}{\omega_{AS}}S_{AS}-\frac{\omega_0}{\omega_S}S_S$ (presented in Fig.~\ref{fig:C}). 
As expected, the cooling condition is satisfied for optical pumping around 203 to 210 THz (1427 nm - 1476 nm). 
This demonstrates that the total phonon energy removal rate can exceed the energy absorbed, leading to net energy removal from the system.

\section{Practical considerations}

There are a few experimental notes that we must discuss in this context.
The escape efficiency of the scattered photons, which is the escape probability of anti-Stokes photons from the material after scattering, must be engineered to approach 100\%. Surface states and band edge states are also important since they create additional uncontrolled scattering and absorption in the material. 
Several different photonic crystal architectures are known to offer the requisite three-dimensional band gaps, including woodpile structure \cite{ho1994photonic, sozuer1994photonic}, inverse-opals \cite{vlasov2001chip} and two-dimensional crystal stacks \cite{joannopoulos2011photonic6}. The selection of a particular crystal design can be made based on our ability to control the aformentioned parasitic effects in each structure.

%
Cooling efficiency, or heat lift per watt of pump power, is a key metric of interest.
For photoluminescence based refrigeration \cite{seletskiy2010laser}, the single-pass cooling efficiency is typically around 0.1\% per interacting (i.e. fully absorbed) photon. In comparison, Raman cooling efficiency is much higher at 5\%-10\% per interacting (i.e. scattered) photon, calculated via the ratio of phonon energy to photon energy. However, since the interaction probability in Raman scattering is small, the single-pass net cooling efficiency is very low around $10^{-7}$ cm$^{-1}$ (Fig.~\ref{fig:C}).
To boost efficiency, photoluminescence coolers employ pump photon recycling through a multi-pass cavity \cite{seletskiy2010laser}. However, due to the high pump absorption which is required for efficient fluorescence, one can only reach a photon recycling factor of about 10, bringing net cooling efficiency into the 1-3\% regime.
In the case of Raman cooling, the cooling efficiency could be boosted up by several orders-of-magnitude into the 1\% regime by adding a high-finesse (Finesse $ \approx 10^5 - 10^6$) resonator around the engineered crystal. This is an advantage of working in the highly transparent regime of the material. Even further efficiency enhancement can come from more careful design of the photonic DoS to express strategically placed anti-Stokes resonances.

\section{Conclusions}

In this study, we demonstrate that the spontaneous Raman cooling of transparent semiconductors is achievable when the photonic DoS of the material is engineered. 
Cooling is reached by significantly enhancing anti-Stokes Raman scattering over the optical absorption probability, along with simultaneous rejection of Stokes scattering events.
As a particular example, we show that our method can be applied to the net laser cooling of undoped silicon, which is currently unattainable by any other laser-cooling method. Furthermore, the cooling of all three triply degenerate modes (1 LO and 2 TO) in silicon is possible simultaneously using a single telecom wavelength pump.

In a broader context, Raman scattering and Brillouin scattering are similar processes differing primarily by the phonon populations they influence. The natural occurence of suitable photonic DoS has already proven to allow Brillouin cooling \cite{bahl2012observation}, thus Raman cooling remains an exciting prospect.
The ability to exert such control on elementary photon-phonon scattering processes through photonic DoS engineering allows us to operate in highly transparent regimes far away from any material absorption, band edges, excitons, or atomic transitions in nearly any semiconductor. Such a broad-based method for laser cooling based on Raman scattering can greatly impact our ability to control the thermal states of matter.

\section*{Acknowledgement}

The authors would like to thank Dr. P. Scott Carney and Dr. Tal Carmon for their encouragement and helpful technical discussions. This work was funded by a University of Illinois Campus Research Board grant (RB15183).

\FloatBarrier

\bibliographystyle{ieeetr}
\bibliography{DoSEngRef}
\end{document}